\documentclass[preprint,aps,prb,showpacs]{revtex4}

\usepackage{graphicx}% Include figure files

\begin{document}

%\preprint{SigmaPhi 2008}

\title{Statistical dynamics of religion evolutions}

\author{M. Ausloos}
\affiliation{%$^1$ 
GRAPES, B5 Sart-Tilman, B-4000 Li\`ege, Euroland 
\\email : marcel.ausloos@ulg.ac.be}
%\affiliation{GRAPES, B5 Sart-Tilman, B-4000 Li\`ege, Belgium}
\author{F. Petroni} 
\affiliation{%$^2$
Universit\'a dell'Aquila, I-67010, L'Aquila, Italy 
\\email : fpetroni@gmail.com}

\date{\today}

\begin{abstract}

A religion affiliation  can be considered as a ''degree of freedom'' of an agent on the human genre network. A  brief review is given  on the state of the art  in data analysis and modelization of religious ``questions'' in order to suggest and if possible initiate  further research, ... after using  a ``statistical physics filter''.
We present a discussion of  the evolution  of 18 so called religions,  as measured through their number of adherents between 1900 and 2000. Some emphasis is made on a few cases presenting a minimum or a maximum in  the  investigated time range, - thereby suggesting a competitive ingredient to be considered, beside the well accepted ``at birth'' attachement effect. 
   The importance of the  ``external field'' is  still stressed through an Avrami late stage crystal growth-like parameter. The observed features and some intuitive interpretations point to opinion based models with vector, rather than scalar, like  agents.
\noindent

\end{abstract}

%\pacs{89.65.Gh, 89.75.Fb, 05.45.Tp}


\maketitle

\section{Preamble}

This paper is based on an invited communication at SigmaPhi-2008, held in Kolymbari, GR. It bears upon previously published work ({\it Europhys. Lett., {\bf 77} (2007) 38002} ) \cite{1} and other materials expectedly to be published \cite{2}. In order to maintain originality
requirements much has been rewritten though the anlayzed data is exactly the same as that used in \cite{2}, i.e. as taken from Barrett  et al. publications \cite{WCT,WCE}. New considerations  and figures are introduced, some of them being even posterior to the Kolymbari presentation. 

\section{Introduction}

Wealth, language, age, sex, ... religion can be considered as degrees of freedom of any agent on the human genre network. In fact, religiosity is one of the most important sociological aspects of human populations \cite{durkheim}.  Like languages and wealth, religions evolve and adapt to the society developments. 

Several questions can be raised, and have been considered from statistical physics points of view, e.g. 
\begin{itemize}
\item 	from a ``macroscopic'' one :  How many religions exist at a given time? 
\item 	from a ``microscopic'' one : How many adherents belong to one religion?  
\item 	does the number of adherents increase or not?
\item  	is there a universal law, whence any model for this ? 
\item 	and why ???  
\end{itemize}

Quantitative answers and mathematical laws have been found or guessed about religious population evolution  \cite{1}. It has been found \cite{1} that empirical laws can be deduced and related to preferential attachment processes, like on an evolving network. However different algorithmic  models reproduce well the data, implying the need for further considerations. A simple growth equation has been shown  to be a plausible one  for the growing evolution dynamics in a continuous time framework. The case of decaying religions is a little bit more complicated.  On one hand, agent based models can be imagined to describe such non-equilibrium processes; on the other hand more formal evolution equations can be proposed. 

%It has been stressed by many that the  empirical data can be very unreliable. 

The Avrami-Kolmogorov differential equation which usually describes solid state transformations, like crystal growth, was used  \cite{2} in order to obtain the preferential attachment parameter introduced previously \cite{1}. It is not often found close to unity, though often corresponding to a smooth evolution. However large values suggest the occurrence of extreme cases which we conjecture are controlled by so called external fields. A few cases indicate the likeliness of a detachment process.   Some cases seem to indicate the lack of reliability of the data.   Two difficulties can be pointed out in the analysis \cite{2}: (i) the  original time of apparition of a religion, (ii) the time of its maximum as measured through the number of adepts or adherents. Both informations are necessary for integrating reliably any evolution equation. However due to the human/sociological information so asked, the questions have no and will remain without a precise answer. Moreover the Avrami evolution equation might be surely improved, in particular, and somewhat obviously, for the decaying religion cases.  Finally, the importance of a so called ``external field influence'' modifying a smooth evolution behaviour has been stressed in several cases \cite{2}.

% Nucleation, growth, aging, death, criticality, self-organization, epidemic spreading, and subsequent avalanches are all keywords well known in statistical physics but also found in sociological systems studies.  
%One of our aims has been  to approach  questions  on religions, through a statistical physics point of view, attempting to quantify religion dynamics as seen from individual adherence distribution functions \cite{1}. 

% \cite{complexity1,complexity2}. 
No need to say that one could consider  several papers as pertaining to the stream we emphasize, whence in some sense we are giving here below the sate of the art of the subject at the time of writing without assertaining that all published results interesting for a statistical physics approach are included.

As usual one could distinguish between algorithmic models from more mechanistic ones. The former ones are at a more complex level than the latter ones, but these have the pertinence of using a few parameters which can be somewhat better controlled.

\begin{itemize}

\item  Surely  the books \cite{WCT,WCE} and paper \cite{futures} appear rather comprehensive in sorting out the various  causes in the ``adsorption/desorption'' process \cite{footnote0,footnote1}

In the same spirit let us mention

\item dynamic models of religious conformity and conversion by Shy \cite{shy}

\item a coevolution model in the book by Dennett \cite{Dennett}

\item and some counterpart on whether religion is due to an evolutionary adaptation  by Dow \cite{dow08}
who observes that most research reports contain agent based simulations  considering religiosity as a mental phenomenon but without  a biological process in which the behaviour is carried by genes.  However bio-cutural theories do exist \cite{dow08,grinde}, including connections with economic considerations \cite{bulbulia}.

\end{itemize}

In the analytical class, i.e. with ``evolution $equations$ content'',  let us point out
\begin{itemize}

\item   a
population growth-death equation has been conjectured  to be   plausible for  the
evolution dynamics in a continuous time framework, i.e.  along the lines of
the  Avrami-Kolmogorov late stage growth equation describing solid state formation \cite{1,2}.

 \item previously, but not known to the authors \cite{1},    Hashemi \cite{hashemi}  has proposed a drift-diffusion model   to express the
time-evolution of density of sect/congregation size

\end{itemize}
 
 Notice that from a 'basic statistical physics point of view''   we have discussed   two different  freely available data
sets on many adherents to religions \cite{1}.   It
 has been found   that empirical laws can be deduced for such a number. Two quite different statistical models were proposed, both reproducing well the data, with the same precision, one being a preferential attachment model % \cite{prefatt} 
 (leading to a log-normal distribution), another based on a ``time of failure'' argument (leading to a Weibull distribution function).

In fact, the tail of the partial distribution function (pdf)  of the number of religions having a given number of adherents was previously presented \cite{zanette}, though hidden in other considerations, unnoticed by many authors \cite{1}. At this zero order statistical level the pdf evolution of Jeowah witnesses was examined  by
 Picoli and Mendes \cite{picoli} who have fitted the pdf with formulae derived in  Tsallis non-extensive statistics.

Among others,  a paper published in Physica A should also retain some attention  where a discussion on    inquisition is  seen in a self-organization mode \cite{roach}, - though one could debate whether the case  is rather not one pertaining the class of ``external field'' influences.

 Leaving aside models of conversion, i.e competition between religions,  arising from theological or economic  selection, 
 should be reconsidered
 with
 attachment and conversion (thus defection) as done in the case of languages. All this {\it in fine} indicates  that a Hamiltonian or Langevinian description  can be of greater interest for the application of statistical physics ideas and techniques to religious population numerical evolutions than  is the case for  languages.
 
 \par The remainder of the paper is organized as follows: in Section II the data bank is briefly discussed, - and criticized, though accepted for further research and subsequent analysis along the theoretical and methodological tools used here  which we adapt to the considered time series set. In Section III the crystal growth-like model \cite{1,2} is briefly recalled. The results are largely presented and discussed in Section IV under the form of Tables and graphs for various   religions, grouping them according to the apparent behaviour and within some ``philosophical class``. Some  discussion and a few concluding remarks are found in Section V and VI. respectively.

\section{Data bank. Theoretical and methodological framework}

Let us quote Iannaccone \cite{ianna98} : {\it Religious data are, on the one 
hand, limited and unreliable. Govern- 
ments collect few religious statistics and 
sponsor little religious research; most 
religious organizations keep sloppy financial records and overly inclusive membership lists; and many aspects of religion are inherently difficult to observe. 
Yet religious data are more abundant 
than most academics realize and far 
more extensive than those pertaining to 
many other ÒnonmarketÓ activities and 
institutions, such as clubs, friendships, 
recreational activities, self-help groups, 
and most social movements. }

She also mentions that {\it 
Institutional records complement 
self-reported survey data. Nearly all denominations track their membership, 
contributions, expenditures, number of 
congregations, and number of clergy, 
and many also keep records on baptisms, conversions, ordinations, missionary activity, and attendance};
% see her footnote8}
 e.g., see Picoli and Mendes \cite{picoli}.

%  The U.S. government collects some relevant data, including statistics on clergy employment and church construction and IRS tax records (which, together with survey data and denominational reports, yield ...
Beginning in 1972, NORCOs General Social Surveys provide (nearly) annual responses to many more religious 
"questions".

Whence data banks indeed exist. However to remain coherent in our own work,  the data \cite{footnote4} so below analyzed here was taken from the World Christian Trends (WCT)  book \cite{WCT}. It is fair to say that this is a remarkable compilation work.
% in table 1-2. 
The tables in \cite{WCT} give information on the number of adherents of the world's main religions and their main denominations: 55 specific  (''large'') religious  groups + atheists + nonreligious, plus a set called other religionists made of 3000 religions
 %which however contains, Yezidis and Mandeans which we consider also, 
so that we have examined 53+2= 55 (truly recognized) religions \cite{2}.  From this data set one has also
information on changes during one century of the number of adherents of
each religion from 1900 till 2000 (information in the data set are given for the following years 1900, 1970, 1990, 1995 and 2000)  -  with a
forecast for 2025 and 2050. %Let us point out that it is not understood (or barely understandable) how such a forecast is made in the data bank.

A critical view of this data has to follow:  we have already \cite{1} noticed a break at $10^7$ adherents, in the pdf,  indicating in our view
an overestimation of adepts/adherents  in the most prominent religions, or a lack of
distinctions between denominations, for these, - as can be easily understood either in terms of propaganda or politics, or because of the difficulty of surveying such cases precisely. Yet one paradoxical surprise stems in the apparent  precision of the data. E.g., in several cases,  the data\cite{WCT} seems to be precise up to the last digit  i.e.,  in mid-2000 , there are  1057328093 and 38977 roman catholics and mandeans respectively.  In strong contrast there are 7000000 and  1650000 wahhabites and black muslims respectively, numbers which are quite well rounded. Thus a mere reading of the numbers  warns about the difficulty of fully trusting the data. Nevertheless the analysis is pursued bearing this $caveat$ here below.  In the following  18 religion evolutions, 
as listed in Table I and II
will be compared. 

\vskip 1cm

 \section{Recall Avrami growth model}
 
We can nevertheless recall that history is full of examples of individuals or entire groups of people
changing their religion, - for various reasons: following the ''leader'',  or ''external pressure'', in face of the martyrdom choice, or  ''internal (social or economic) pressure''  or so called adaptation under proselytism action. ''Competition'' through interactions or under ''external field conditions'' exist in many cases. In this way, the number of adherents can  evolve drastically due to such various conditions \cite{roach}. External field conditions are rather more  drastic  and frequent in the religious domain than in e.g. language history.
%See also Appendix A  in \cite{2} for some discussion outlining   aspects, like ''differences'' between languages and religions, from a physics point of view, perspective or input into modeling such sociological features.

%yet  recognized that thre is rarely  a religon  without a church.

We recognize that the definition of a religion \cite{dow07} or an adherent (or adept) might not be accepted univocally.
One might   distinguish between adepts and adherents, as well as debate on the ''intensity'' of the adhesion or measure the ''importance'' of a religion thorugh other indicators \cite{Herteliu}.  This ''intensity'' of a sociological attitude is a well known problem in sociological studies. It cannot be avoided. However our primary goal is not discuss data acquisition on religious adherence. We will take for granted WCT published data, recognizing that we have as much as other authors, including the data surveyors,  some doubt about the exactness and validity of the data. 

Nevertheless one can expect  to more precisely certify the religious adherence of an agent than e.g. the linguistic one.   Indeed one can hardly be multi-religious but one can be a polyglot.  
Whence   the fundamentally relevant variable  is taken here below as the number of adherents of a ''religion'', under ''usually accepted'' denominations. 

Of course one can also switch more easily, i.e. through ''conversion''  from one religious denomination to another than in language cases. Thus the observation time of a religious state needs very careful attention in surveys. In the same framework,  the  time axis, it is hard to know precisely when a religion was born. The origin is usually quite conventional.  The time life, or aging, of a religion can be studied through the number of adherents, surely for modern times, but with some uncertainty on the initial conditions, here the initial time.

\begin{figure}
\centering
\includegraphics[height=25cm,width=15cm]{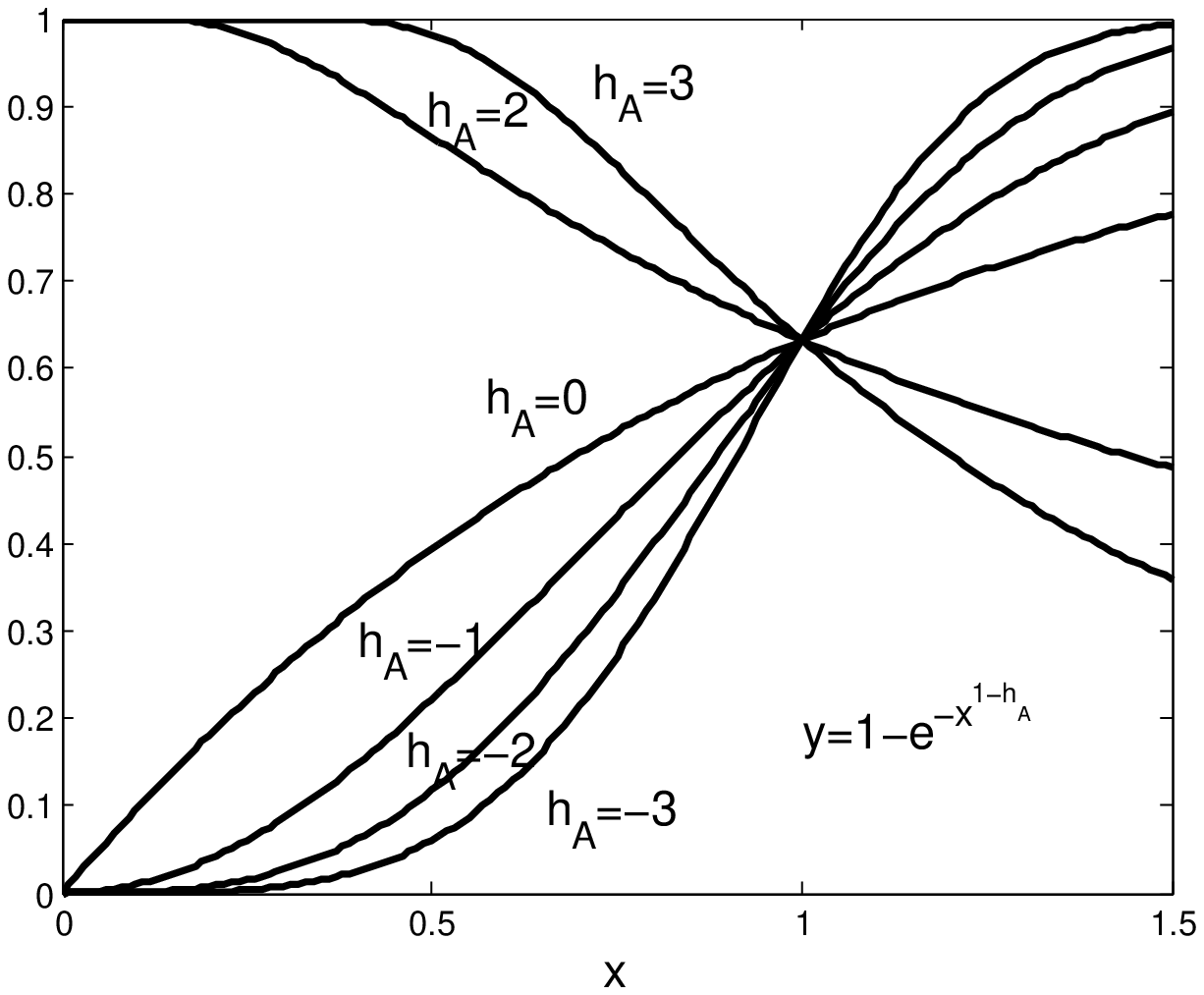}
\caption{\label{fig1bA}  Theoretical behavior of the solution of an Avrami Equation as (Eq.(5) in reduced units for various typical $h_A$ values}
\end{figure}

The
population growth-death equation   conjectured in \cite{1} is  a first approximation plausible modeling of the
evolution dynamics in a continuous time framework, i.e. the time evolution of several ''main'' religions, from a microscopic interpretation point of view looks like
the growth Avrami-Kolmogorov equation, describing solid state formation in a continuous time framework. The solution of which is usually written as
 
\begin{equation}\label{Avrami}
f(t)=1- exp[- K t^{n} ]
\end{equation}
where $f(t)$ is the volume fraction being transformed from one phase to another; $K$  and $n$ are adjustable parameters (Fig. \ref{fig1bA});
this Avrami equation is of interest for so called late stage growth, i.e.
\begin{equation}\label{logisticmap}
\hat f(t)=\frac{1}{1+exp[- K t ]}
\end{equation}

 {\it A priori} in analogy with crystal growth studies \cite{auslooscrystalgrowth,Gadom}, we have considered that a microscopic-like, continuous time differential equation can be written for the evolution of the number of adherents, in terms of the percentage with respect to the world population,  of the world main religions, as for competing phase entities in Avrami sense
\begin{equation}\label{Avramidiff}
{d \over dt} g(t)=\gamma t^{-h_A} [1-g(t)].
\end{equation}
It can be solved easily giving the evolution equation for the fraction $g(t)$ of religion adherents  
\begin{equation}\label{Avrami_sol}
g(t)=1-\eta  \;  exp\left[{-\frac{\displaystyle \gamma}{\displaystyle1-h_A} \displaystyle t^{1-h_A}}\right]
\end{equation}
where, adapting to our case this 
Eq. (4), $\eta$ is related to the initial condition, $\gamma$ is a (positive for growth process) rate (or scaling) parameter to be determined, and $h_A$ is a parameter  to be
deduced in each case, measuring the attachment-growth (or death) process
in this continuous time approximation. 

% The {\it relaxation time} $\tau_n$,  since $n\equiv 1-h$, of this stretched exponential growth is 
%\begin{equation}\label{tau_n}
%\tau_n = \left(\frac{\gamma}{1-h}\right)^{-1/(1-h)}
%\end{equation}
%which is markedly rate ($K$) dependent.

For further consideration, let us explicitly write the ''very (infinitely) slow'' growth case $h_A$=1, i.e., 
\begin{equation}\label{Avramidiffh1}
{d \over dt} g(t)=\gamma t^{-1} [1-g(t)],
\end{equation}
whence 

\begin{equation}\label{Avrami_sol1}
g(t)=1- \beta t^{-\gamma},
\end{equation}
where $\beta$, being positive (negative) for a growth (decay) case, is set by initial conditions; for $h_A=1$,  there is no ''relaxation time'', but a scaling time $\tau_1$ = $\beta^{1/\gamma}$, or $\beta = \tau_1^\gamma$.

\begin{figure}
\centering
\includegraphics[height=25cm,width=15cm]{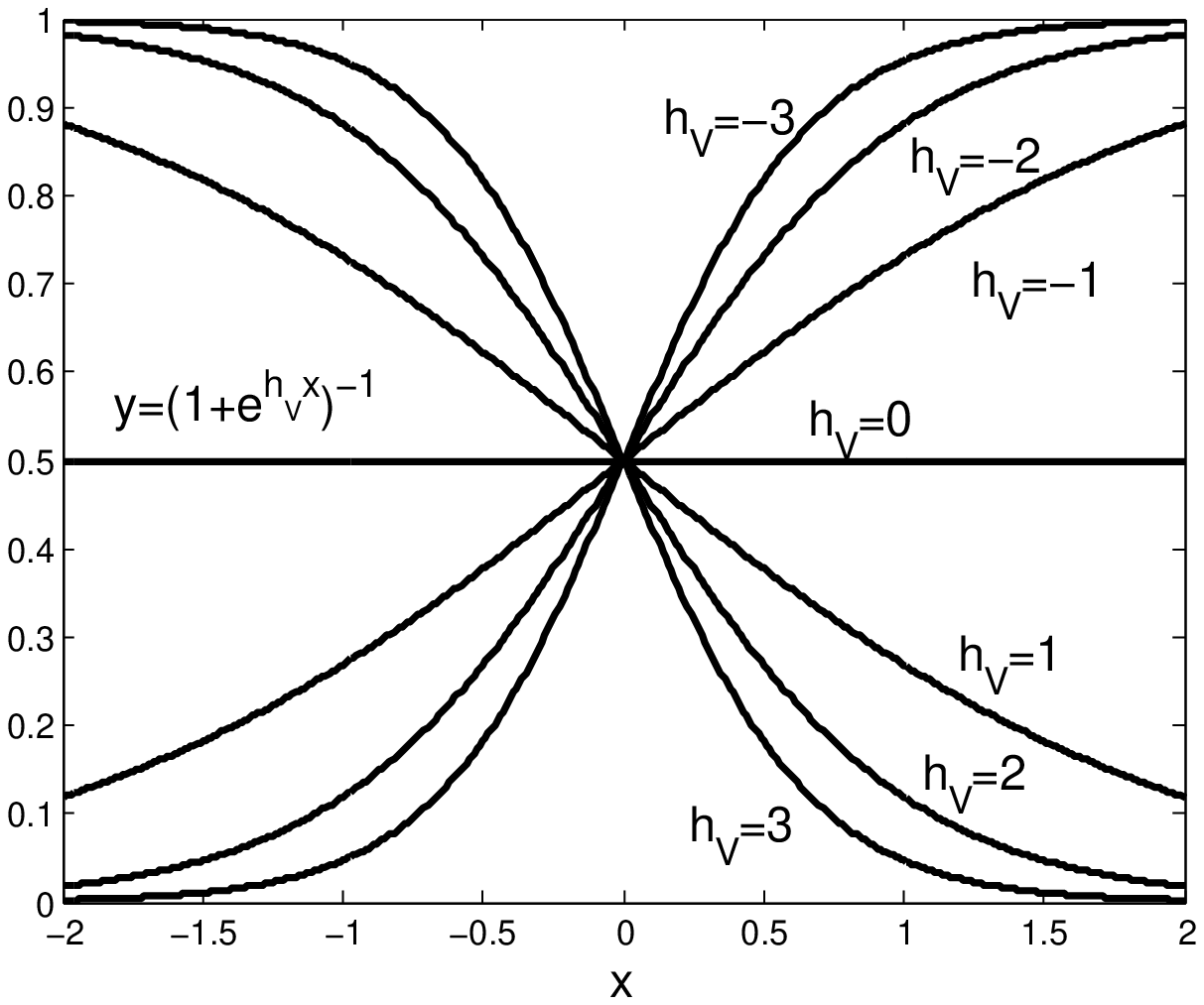} 
\caption{\label{fig1aV}   Logistic map or Verhulst  law (Eq.(2))   in reduced units for various typical $h_V$ values}
\end{figure}

The $h_A$-cases which can be illustrated through an Avrami equation are shown in arbitrary time units in Fig.1 for various $h_A$ values, for $\eta$ =1 and $\gamma= 1-h_A$. They are compared to the (generalized, i.e. $n \neq 1$) logistic map (Fig.2).

What should be emphasized is the fact that religions have appeared at some time $t_0$ which is somewhat unknown, or to say the least, loosely defined, due to a lack of historical facts but also due to the inherent process of the creation of a religion. 
Yet this initial time is a parameter much more important than in crystal growth, but less well defined. Therefore we rewrite the Avrami equation as
 
\begin{equation}\label{Avrami_t_0}
g(t)=1-exp\left[{-\left({t-t_0 \over t_1}\right)^{1-h_A}}\right],
\end{equation}
thereby allowing also for a time scaling through $t_1$ related to some growth (or death) rate process. Notice or so that the maximum in such theoretical laws occurs at zero or +/- infinity, - a time information on which there is not much data in the case of religions. 

When  the number of adherents  presents a maximum or a minimum which can be recognized to occur during the present or a recent century we will use 
a second order polynomial  like $y=A+Bt+Ct^2$ for the fit.

If $(t-t_0)/t_1$ is much smaller than 1,  Eq. (\ref{Avrami_t_0}) can be expanded in Taylor series, taking only the first order, and gives
\begin{equation}\label{Avrami_sol2}
g(t)=\alpha+\left({t\over t_1}\right)^{1-h_A}
\end{equation}
where we have mapped the time axis onto an $x$ axis starting from $0$ (instead of 1900,  being this completely arbitrary and purely conventional) and $\alpha$ representing the initial condition, i.e. the value of the number of adherents for $t=0$.  This is obviously the most simple non linear expression, i.e. 3 unavoidable parameters, which can be used.

 Sometimes it is readily observed from the WCT tables that there are either ''presently growing'' or "presently decaying"  religions, for which a minimum or after some recent maximum is observed during the 20-th century.  For such ''religions''   the number of adherents  can be in a first approximation fitted with
a second order polynomial  $y=A+Bx+Cx^2$ for which the parameters are 
given in  Table \ref{table3}), without  presently having  a modelization of such a behaviour.

\section{Results}
 
 The number of adherents has been transformed as relative proportion, thus in per cent (pc) of the supposed world population during the surveying year(s). We recognize  the lack of precision of such a value. In all cases the least-square best fit has been made over 5 points, represented by dark  symbols  in the figures. The WCT forecasts are  indicated by an open but corresponding symbol with a cross inside.  The fits  are usually rather good, but we emphasize that the extrapolation either often  overshoots  or sometimes underestimates the WCT forecast for 2025 and thereafter, except for  cases cases for which $0 \ge h \ge -1$.  This disagreement is partially due to the lack of saturation implied by the Avrami model.

Results of the $h_A$-fit to Avrami equation of the WCT surveys \cite{WCT} are summarized in Tables \ref{table1} and \ref{table3}.
The parameter $h_A$ value  and  meaning deserve some short explanation and discussion. According to the standard growth (Avrami) process $h_A$ should be positive and less than 1,  since $n\equiv 1-h_A$;  if it is
greater than 1, this is indicating the possibility for $detachment$.  We consider that if $|h_A|$ is outside  the $(0,1)$ interval, we have to imagine that the nucleation growth process is heterogeneous and/or conjecture that it is due to {\it external field} influences. Moreover notice that when $h_A$ is greater than 1, the Avrami equation solution decays, ... from a maximum at the time $t_0$. However it is hardly difficult to know when a religion has attained its maximum number of adherents.  Thus the time scale or the initial appearance time of a religion are questionable points. %Another point is obvious from Fig.1. %The theoretical expressions do not allow a fit in the vicinity of either a maximum or a minimum. %We should expect deviations, if such a case occurs, whence other empirical functions to be of interest.%

In \cite{1}  the main denominations were ''loosely grouped''. To be more specific: Christians in \cite{1} were the results of grouping together 12 denominations; similarly for Muslims we grouped 15 denominations.   A  somewhat more detailed analysis on 58 ''time series''  can be found in \cite{2}, as taken from the  World Christian
Encyclopedia (WCE) and World Christian
Trends (WCT) reference books \cite{WCT,WCE}.  Let us examine several cases according to their loosely defined "connections" in a philosophical sense and present comparative results for various groups, starting from the less populated ones, approximately at most 0.001  to the biggest ones, 0.20   at most.

%\newpage
\begin{table}\caption{Values of the parameters $h_A$
%$\alpha$, 
and $ t_1$,  used for fitting the data of ''increasing religions'' with a power law formula; see Eq. (\ref{Avrami_sol2}); religions are hereby  ranked based on the size of the attachment parameter $h_A$ which can be negative or positive but $\le 1$,  and on ''decreasing religions'' with Eq.(\ref{Avrami_t_0}); $h_A$ is in this case $\ge 1$ }
\begin{center} \begin{tabular}{|l|r|r|} 
\hline Religion & $h_A$  & $t_1$  \\ \hline 
%Shaivites & -5.32 & 0.032 & 239 &O&2\\ 
%Hanbalites & -4.66 & 0.000305 & 527&O&3 \\ 
%Hanafites & -3.84 & 0.0629 & 211&O&2 \\ 
Zoroastrians & -3.64   & 530  \\ 
%Kharijites & -2.88 & 0.000196 & 1150&O&3 \\ 
Afro-Caribbean religionists & -2.75 &  1800   \\ 
Black Muslims & -2.36 & 1110  \\ 
Pentecostals/Charismatics & -2.19  & 208  \\ 
Independents & -1.61 & 288  \\ 
%Shafiites & -1.49 & 0.024 & 528&O&2 \\ 
Afro-American spiritists & -1.32 & 4990 \\ 
%Ithna-Asharis & -1.26 & 0.0137 & 812&U&4 \\ 
Afro-Brazilian cultists & -1.21  & 2610  \\ 
%Zaydis & -1.10 & 0.000741 & 3450&=&4 \\ 
%Alawites & -1.09 & 0.000154 & 7560&=&4\\ 
%Ismailis & -1.04 & 0.00142 & 1870 &O&4\\ 
%Yezidis & -1.01 & 1.84e-005 & 2.23e+004&O&4 \\ 
%High Spiritists & -0.83 & 2.33e-005 & 5690 &O&3\\ 
%Sikhs & -0.792 & 0.00182 & 3130&O&3\\ 
%Ahmadis & -0.789 & 4.32e-005 & 4170&=&4 \\ 
Baha$'$is & -0.368 & 1.38e+004 \\ 
%Druzes & -0.366 & 4.38e-005 & 8.85e+004&=&4 \\ 
%Neo-Hindus & -0.212 & 6.19e-005 & 1.28e+004&O&2 \\ 
Marginal Christians & -0.206  & 1e+004  \\ 
%Mandeans & -0.0667 & 5e-006 & 3.17e+007&U&5 \\ 
%Malikites & 0.0566 & 0.0167 & 63803 &=&5\\ 
%?Other sectarian Muslims & 0.0929 & 2.62e+006 \\ 
crypto-Christians & 0.230  & 1.8e+004  \\ 
%Reform Hindus & 0.384 & 0.000154 & 1.78e+007&O&6 \\
%?Chinese folk-religionists & 1.07  & 3.49e-015  \\ 
Orthodox & 1.14  & 1.06e-008  \\ 
\hline\end{tabular} \end{center} \label{table1}
 \end{table}

%\newpage
\begin{table}\caption{Values of the parameter used for fitting data on 4 ''decreasing'' and 3 ''increasing''  religions with the polynomial equation  $Cx^2+Bx+A$}
\begin{center} \begin{tabular}{|l|r|r|r|} 
\hline Religion & $C$ &  $B$ & $A$  \\ \hline 
Nonreligious & -2.61e-005 & 0.103 & -102  \\ 
Atheists & -1.31e-005 & 0.0514 & -50.3  \\ 
unaffiliated Christians & -4.38e-006 & 0.017 & -16.5 \\ 
Roman Catholics & -4.2e-006 & 0.0165 & -16  \\ 
%New-Religionists (Neoreligionists) & -3.88e-006 & 0.0153 & -15&U&12\\ 
%Shamanists & -4.87e-007 & 0.00185 & -1.74&U&12 \\ 
%Confucianists & -2.11e-007 & 0.000831 & -0.815&U&12 \\ 
%Wahhabites & -5.53e-008 & 0.000215 & -0.208 &U&12\\ 
%Taoists & -4.4e-008 & 0.000174 & -0.171&U&12 \\ 
%???Other religionists (in 3000 religions) & -3.81e-008 & 0.00015 & -0.148  \\ 
%Ashkenazis & -2.46e-008 & 4.26e-005 & 0.0149&U&11 \\ 
%Oriental Jews & -3.23e-009 & 1.22e-005 & -0.0112 &O&11\\ 
\hline
%Samaritans & 7.14e-012 & -3.01e-008 & 3.18e-005&U&14 \\ 
%Sefardis & 6.52e-010 & -2.8e-006 & 0.00315 &O&14\\ 
%Jains & 8.97e-009 & -3.61e-005 & 0.0371&U&14 \\ 
%Shintoists & 2.05e-007 & -0.000836 & 0.853&=&14 \\ 
%Saktists & 2.12e-007 & -0.000824 & 0.806 &O&9\\ 
Protestants & 7.66e-007 & -0.00306 & 3.11  \\ 
Anglicans & 9.75e-007 & -0.00386 & 3.83  \\ 
%Vaishnavites & 1.25e-006 & -0.00487 & 4.82 &O&9\\ 
%Sufis & 2.34e-006 & -0.00921 & 9.11&=&10 \\ 
%????Animists & 2.77e-006 & -0.0111 & 11.2  \\ 
Evangelicals & 5.95e-006 & -0.0233 & 22.8  \\ 
\hline \end{tabular} \end{center} \label{table3} \end{table}

%increasing in  Islam, which is known to be to-day the most important denomination. decaying (e.g., Ethnoreligions) or rather stable (e.g.,  Christianity and Buddhism) is already shown in Fig. 4 of  \cite{1}. 

 \begin{figure}
\centering
\includegraphics[height=25cm,width=15cm]{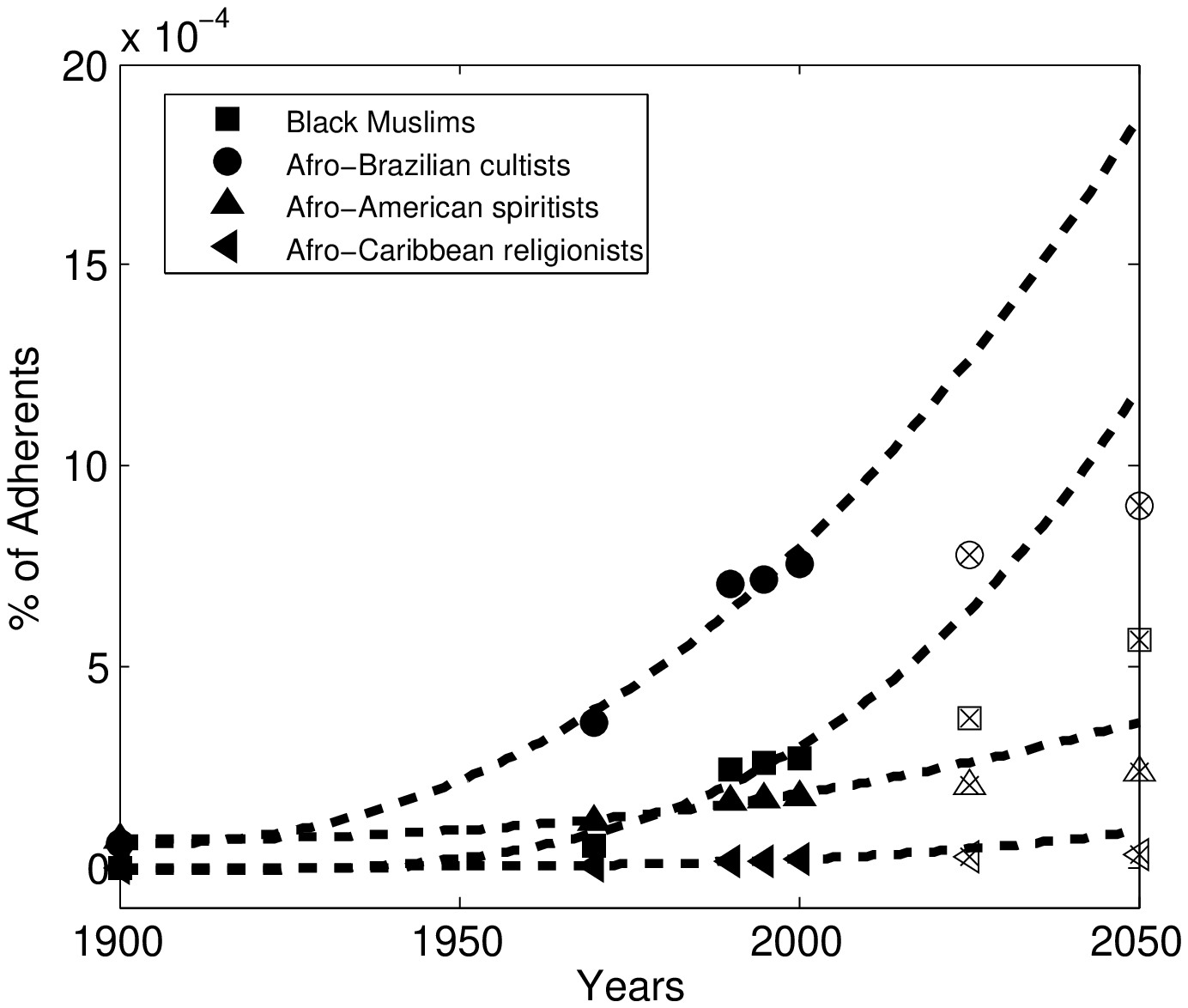}
\caption{\label{Fig3black}   Evolution of the relative to the world population number of adherents in four small denominations, locally concentrated and rather new}
\end{figure}

\begin{figure}
\centering
\includegraphics[height=25cm,width=15cm]{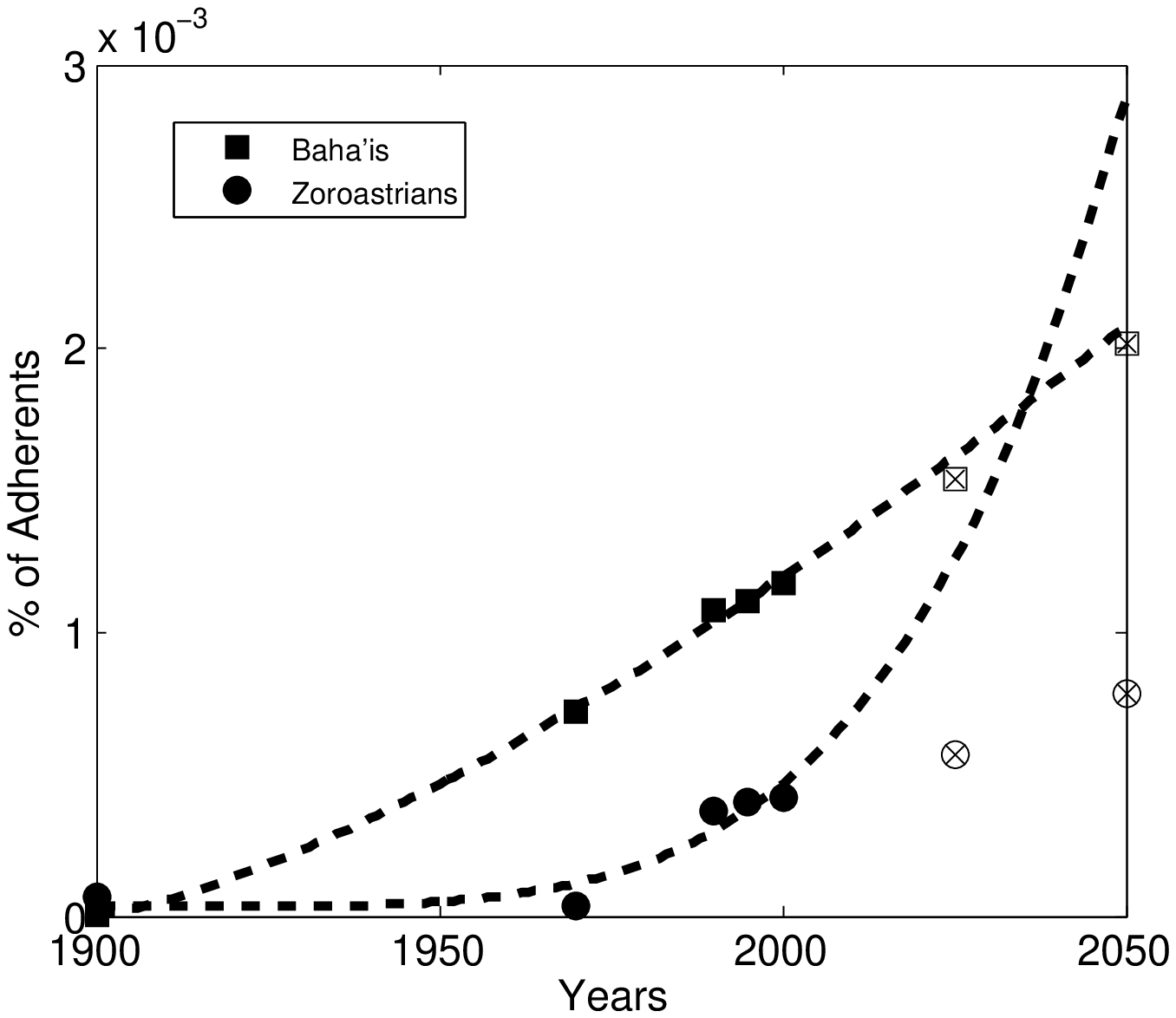}
\caption{\label{fig4baha} Evolution of the relative to the world population number of adherents in two small denominations, locally concentrated and rather discriminated}
\end{figure} 

\begin{figure}
\centering
\includegraphics[height=25cm,width=15cm]{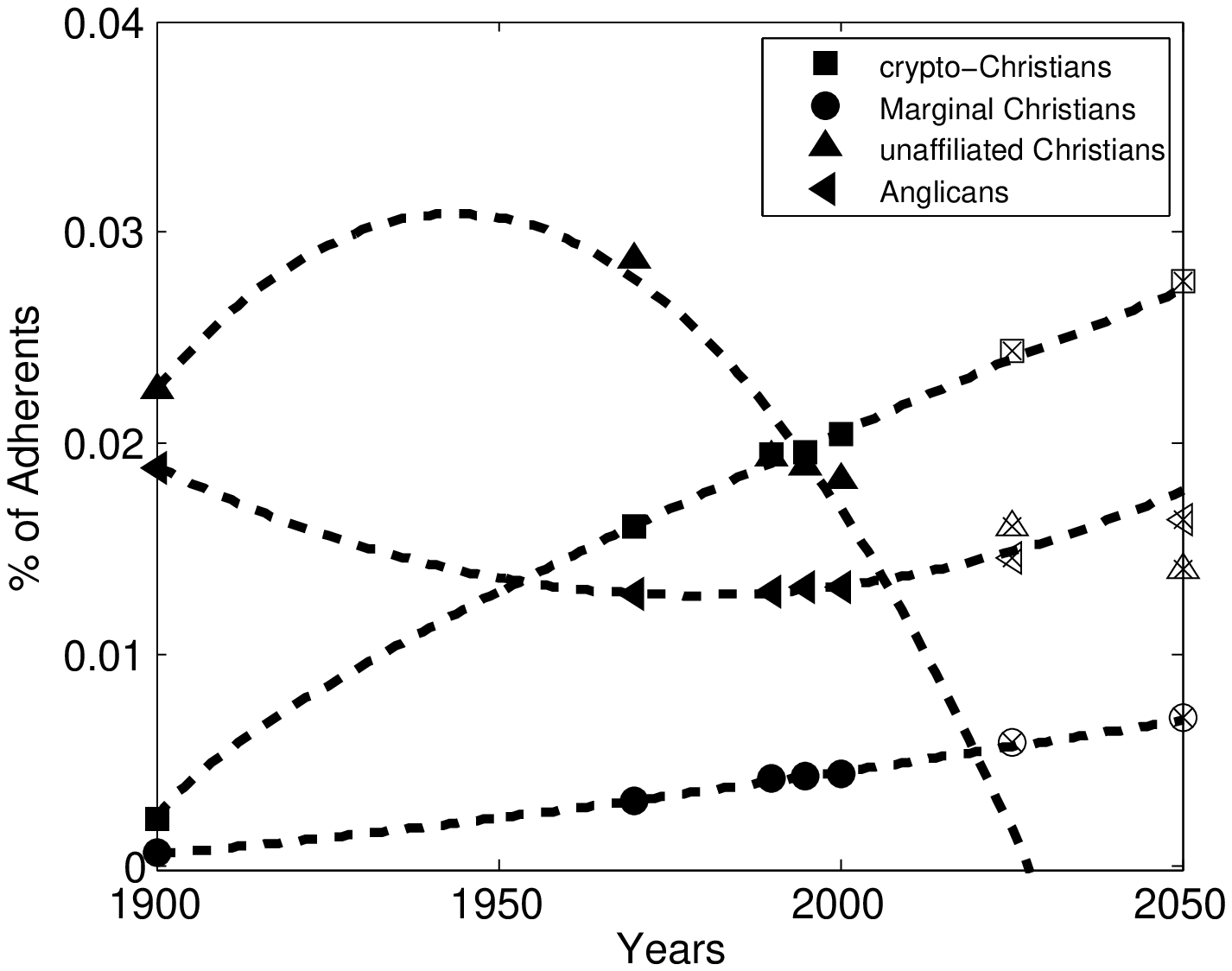}
\caption{\label{Fig5christ}   Evolution of the relative to the world population number of adherents in four large denominations, related to christian beliefss}
\end{figure}

 \begin{figure}
\centering
\includegraphics[height=25cm,width=15cm]{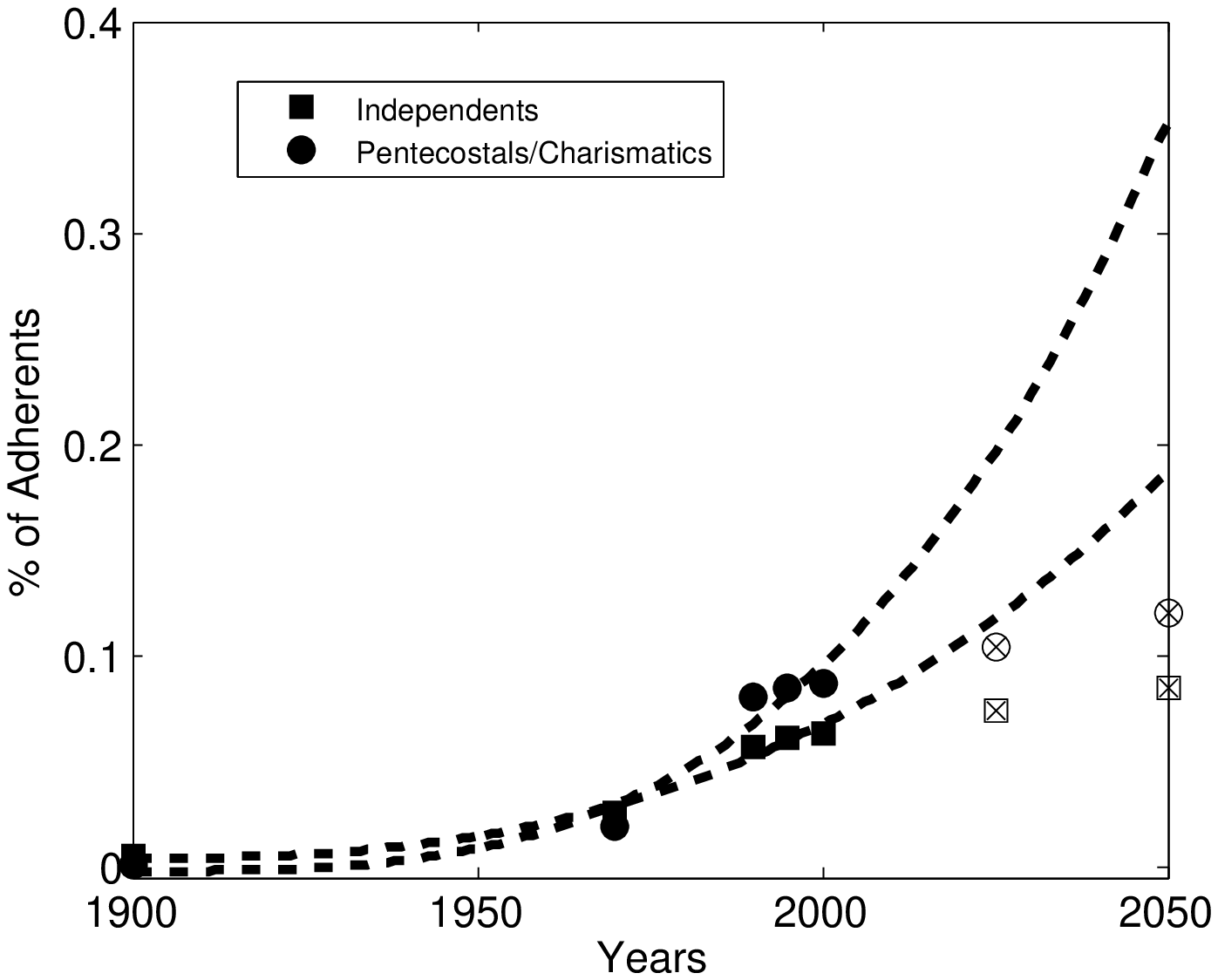}
\caption{\label{Fig6charism}  Evolution of the relative to the world population number of adherents in two    large denominations  related to christian beliefs}
\end{figure}

\begin{figure}
\centering
\includegraphics[height=25cm,width=15cm]{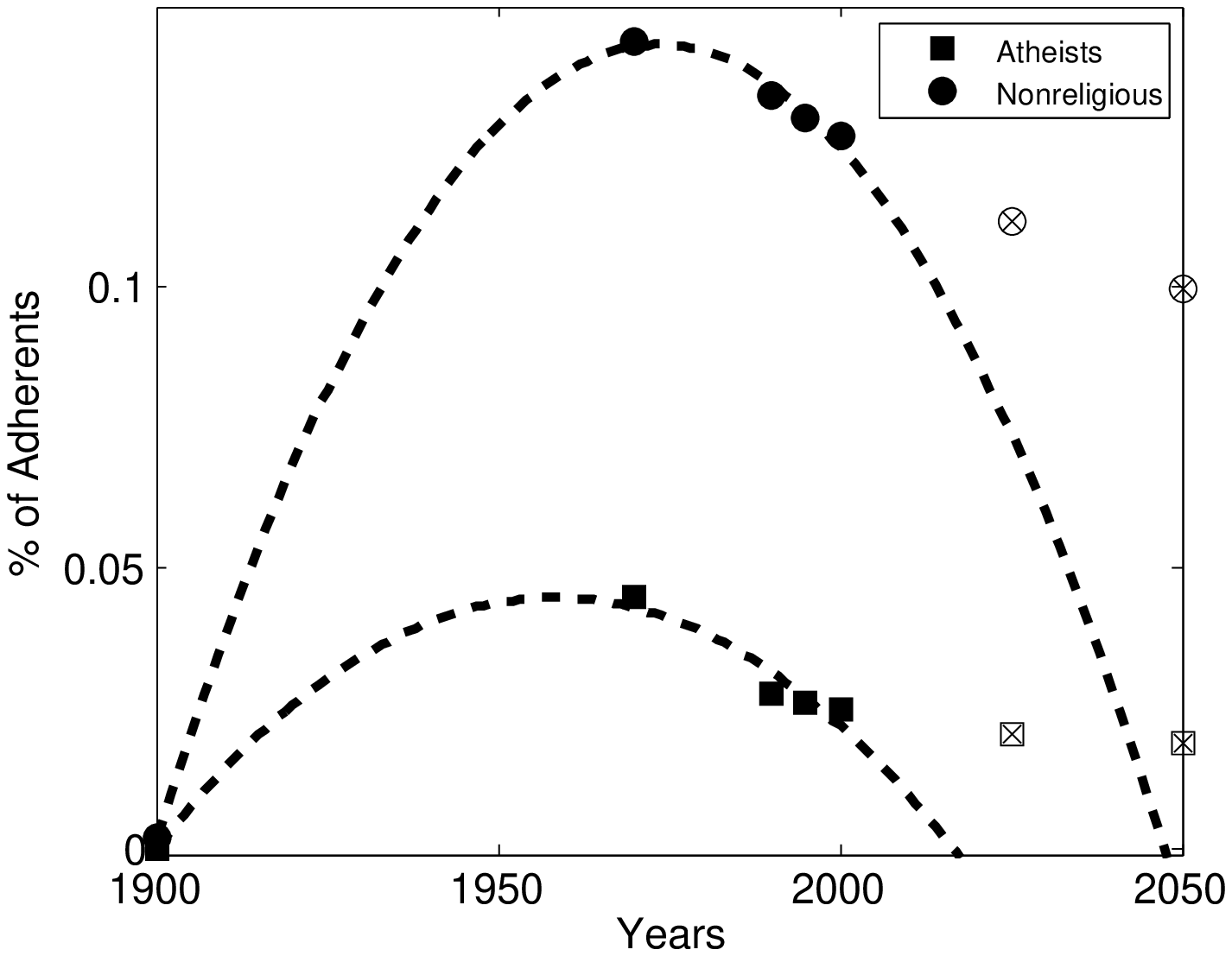}
\caption{\label{Fig7ath}  Evolution of the relative to the world population number of adherents in two   groups outside religious beliefs }
\end{figure} 

 \begin{figure}
\centering
\includegraphics[height=25cm,width=15cm]{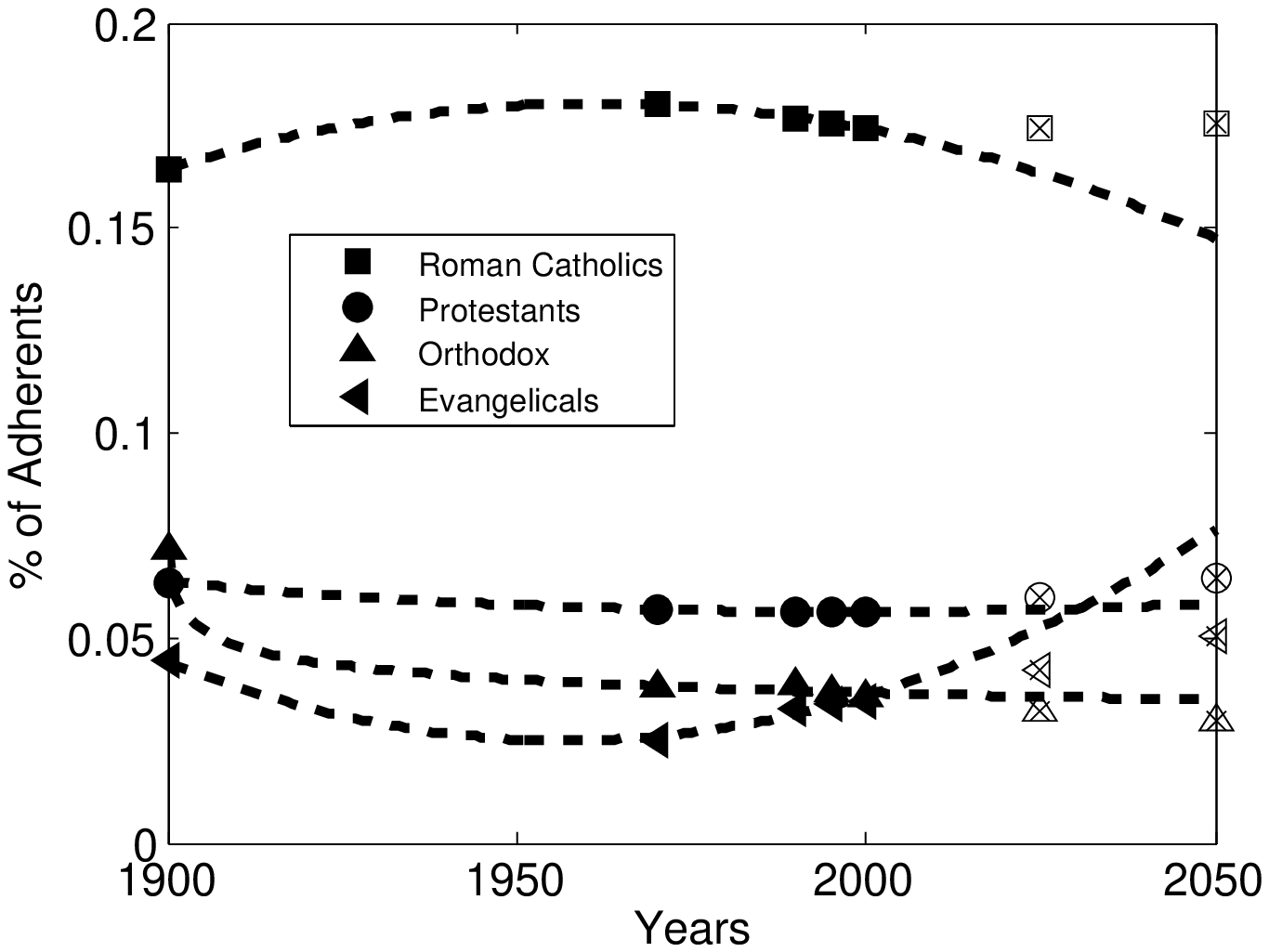}
\caption{\label{Fig8catho}    Evolution of the relative to the world population number of adherents in the four largest denominations related to christian beliefs}
\end{figure}

 In Fig. 3, a set of black  christian and islam based related religions is shown.  Their growth is pretty obvious. The geographical  localization of such groups is  rather limited, and the socio-economic level of the adherents usually not the highest one. The $h_A$ parameter is near -2, and the scaling time around 2000.  We may here recall the ''high'' growth  also seen \cite{2} for  Hanfites, Shafiites and Malikites  which are all Sunnis. 
 
 This should be compared or contrasted to the cases of Fig.4 for Bahais and Zoroastrians. Both are "not politically" nor "religiously" approved by the main rival religions or the political regime(s).  The former has a steady growth. the latter  is a very old religion,  officially replaced by more modern ones, like  Christianism or Islam, with adepts mainly based in Iran, but who are not supposed to be "recognized". One may debate on the data value, i.e.   the case of Zoroastrians (in Fig. 4) indicates an anomalous point corresponding to 1975. One might recall a similar religious attitude masking a political one, in Poland during the communist time.  Religion being a mask for an ndentification to a political (in both cases somewhat illegal) opinion. Maybe we should not need to add a comment based on ''political considerations'' here, but we may consider that the meaning of $h_A$ makes sense again.

The case of a few christian based  "small" denominations, but more widely spread than in Fig.3,  is illustrated in Fig.5. Two are rather steadily growing, but anglicans have been decaying during the last century and are now re-growing,  while unaffiliated christians are  quickly disappearing after a maximum in the 1940-1950
range. Although one is a  growing and the other a decaying denomination, within a similar philosophical origin, one should not immediately conclude that  there is a transfer from one to the other. Yet this different behaviour points to the interest of measuring the religious transfers, if any,  as in language studies.

Though small in membership, such an ensemble of ``sects'' and ``cults'' are prominently  featured in media stories, public debates, and legal disputes about the place of religion in society. One encounters repeated claims that participation in such groups should not be viewed as the exercise of religious freedom but rather as enslavement to organizations bent on ``brainwashing'' and exploitation.Thus it seems natural to expect some high "volatility" in the number of adherents, and the number of religious denominations as well. 

 With respect to the cases presented in Figs. 4-5, one should  observe the fast growth of independents and charismatics in Fig. 6 which have $h_A$ close to -2, but $t_1$ quite small ($\simeq 250$). Further research on the christian varieties and adherence volatility, beside transfer intra and inter christian denominations,  seems of interest.

Although they are not truly religious denominations,  atheists and religionists are part of the "religious world" also. It might at first appear surprising that these two groups  reach a maximum in the 1950's and are now much decaying (Fig.7). One fundamental question related to the above comment is : where are they going ? ... !  Some simulation based on "ageing" might be derived {\it mutatis muntandis} in adapting Shya \cite{shya} simulation with explanation on changes in the 
proportion of secular and religious people.

Finally,  the main christian based denominations are examined in Fig. 8. As in Fig. 5 for smaller denominations, growth after a minimum  (evangelicals) and decay after a maximum (roman catholics) can be observed, with two (protestants and orthodoxes)  decaying denominations. The similitudes and contrasts between the two figures are amazing. Can we suppose that indeed a simple model of transfers will be workable and reproduce such features?

 % One can debate whether the parabola makes physical sense of course, and whether it is a too rough approximation of the Avrami (or logistic map) analytical form in a too limited  time domain.

\section{Discussion}

We consider that the religious practice is more likely more diverse than WCE and WCT surveys indicate. 
We indicate that with an Avrami late stage  crystal growth-like equation equation the fits to the relative adherence evolution can be quite often good, in particular for the growing cases.  Physically speaking that gives some support to the conjecture of  religions growing like crystals \cite{iop}. However we cannot expect that the Avrami equation holds true for ever; the system should saturate at some point, except if only a few religions are excessively predominating, and not ''allowing'' the probability of existence (in a thermodynamic sense) of others. 
Indeed most of the fits seem often to indicate  an exponential behavior. This is clearly wrong on the long term. Whence it would be of interest to develop an alternative set of fits and considerations through the logistic map approach, recalled in the introduction (Fig.2). In this way the growth of the population due to the inherently limited resources would be more realistically taken into account. However this introduces an extra  social parameter, hard to measure or conceive.

The same is true for the parabolic fits, which either indicate  a quite quickly forthcoming disappearance of a religion or allows for infinite growth.  We recognize that these are approximations.  Of course one could remap the parabolae into the exponential expansions, allowing for error bars in the identification of coefficients, but this has not been done yet.

Last but not least, it might considered that the growth in many cases only reflect the population growth in several countries. %\cite{JJSDS}
   It is true that it would be interesting to recount the number of adepts, and their rate of adherence, per religion $and$ per country, and to correlate the evolution to the birth rate in the examined  country with the religion(s) growth rate, and in particular the attachment parameter value. The latter might not be a constant as assumed here above.

Finally, it could be of interest to obtain and analyze data taking into account the emergence of denominations, rather than their long time behaviour.   Let us point to a
  dramatic   case in Japan after
World War II, when the abolition of state Shinto and advent of religious freedom led to a five-year period 
known as The Rush-Hour of the Gods 
during which some 2,000 new sects and 
cults were formed. See also Upal \cite{upal} on the emergence of new religious movements. 
Shoud not this be an interesting system with quite a different time scale, with heterogeneous or not fluctuations,  thus of interest in statistical physics?

\section{Conclusion}

In conclusion, here above we have shown that we can attempt to make a statistical physics like analysis of the number of adherents in religions, going beyond our first paper \cite{1} on the subject. However the data seem sometimes barely reliable. Nevertheless one can, expecting better surveys, at a more limited scale, suggest further lines of research.

Rates of growth and decay are surely not the only 
characteristics that vary across denominations \cite{ianna98} nor should they only be taken without other measurements and out of context. The time dependence of the  number of adherents can be considered to be a very restrictive    way to ''measure''  the evolution of a religion.   One could also ''weight'' the level of adherence to a religion. For example, one could try as for languages to define a religion through its quantity of practitioners, rituals,  ....  Many other indicators are possible \cite{footnotestatus}. One can measure diverse quantities related to the  religious effect.  Yet we observe some general features. More modelization and simulation are still in order beside those already published \cite{futures}. We suggest to let religious adherence to be a degree of freedom of a population, and take it through statistical physics considerations for our enlightment.  
Models of opinion formation are obviously in order.
One could suggest agent based models like for languages, including the role of external fields. 
 Sex, age, memory,  location, environment considerations,  are to be taken into account, whence heterogeneous agents with vector-like properties. Correlations to other socio-economic features will allow a more interesting qualification of models.
 
One could try to have a Langevin equation connexion to Avrami equation;  of course  we need to define a hamiltonian $H$ and a current: that implies interactions thus competitions between entities; what we do not see here yet.   However the hamiltonian can be obtained following standard  ideas, like turning over the pdf into its log and defining some temperature. Religions seem to be an interesting field of study for statistical mechanics!

                \acknowledgments
The work by FP was originally supported by European Commission Project
E2C2 FP6-2003-NEST-Path-012975  Extreme Events: Causes and Consequences.
Moreover this paper would not have its form nor content without comments and constructive criticisms by many coworkers, the list being too long to write, see 	 a very selected  few in \cite{1,2} and most of all the A. Carbone and G. Kaniadakis invitation to  present such considerations at SigmaPhi 2008.


\begin{thebibliography}{99}
\bibitem{footnotestatus}         It is indeed clear that a religious adherent instead of being an analog of an up or down spin, is rather a vector for which each element can be a quantity measuring some value like one of those considered  in sociology, i.e. a ''quality''. The adept attaining an extreme value of one or several vector components.  Next we may imagine Potts vector or ferrroelectric type (Hamiltonian) models for describing an ensemble of religious  agent state or evolution.
\end{thebibliography}
\end{document}